%% file: main.tex
\documentclass{article}
\usepackage{spconf,amsmath,graphicx}
\usepackage{colortbl}
\usepackage{mathtools}
\usepackage{amsthm}
\usepackage{amsfonts,amscd,keyval}
\usepackage{bm}
\usepackage{multirow}
\usepackage{dsfont}
\usepackage{float}
\usepackage{xspace}
\usepackage{bbding}
\usepackage{graphics}
 \usepackage{enumitem}
 \usepackage{breqn}
\usepackage{booktabs}
\usepackage{xcolor}
\usepackage{hyperref}

\title{Parameter-Efficient Transfer Learning Under Federated Learning For Automatic Speech Recognition}
\name{{Xuan Kan$^{\dagger*}\thanks{*Corresponding author: Xuan Kan $<$xuan.kan@emory.edu$>$. This work was done while Xuan Kan was an intern at Google.}$ \quad Yonghui Xiao$^{\ddagger}$ \quad  Tien-Ju Yang$^{\ddagger}$ \quad  Nanxin Chen$^{\ddagger}$\quad  Rajiv Mathews$^{\ddagger}$}}
\address{$^{\dagger}$ Emory University, 
    $^{\ddagger}$ Google LLC}
\begin{document}
%
\maketitle
\begin{abstract}
This work explores the challenge of enhancing Automatic Speech Recognition (ASR) model performance across various user-specific domains while preserving user data privacy. We employ federated learning and parameter-efficient domain adaptation methods to solve the (1) massive data requirement of ASR models from user-specific scenarios and (2) the substantial communication cost between servers and clients during federated learning. We demonstrate that when equipped with proper adapters, ASR models under federated tuning can achieve similar performance compared with centralized tuning ones, thus providing a potential direction for future privacy-preserved ASR services. Besides, we investigate the efficiency of different adapters and adapter incorporation strategies under the federated learning setting.
\end{abstract}
\input{sections/1-intro.tex}

\input{sections/2-method.tex}
\input{sections/3-exp.tex}
\input{sections/4-con.tex}



\vfill\pagebreak


\bibliographystyle{IEEEbib}
\bibliography{refs}

\end{document}

%% file: sections/1-intro.tex
\section{Introduction}

With the rapid development of Large Language Models such as Bard and ChatGPT, computers can now possess near-human-level competency in understanding language, enabling human-computer interactions using natural language. This advancement has positioned Automatic Speech Recognition (ASR) as a necessary component in future human-computer interfaces, making it a common element in smart devices and accentuating the demand for efficient ASR services. Current cutting-edge ASR services are built on giant deep neural networks that consist of a huge number of parameters and require extensive data for training~\cite{gulati2020conformer, conneau2022fleurs, radford2022robust}. Yet, they fail to deliver flawless performance across various scenarios. Transfer learning with domain-specific user data can address this issue~\cite{52517}, but ASR service providers cannot gather such data from users due to the higher sensitivity of voice data than other forms of data. Thus, improving model performance for user-specific scenarios while ensuring data privacy presents a significant challenge. 

Federated learning is a promising solution for balancing data privacy concerns and the extensive data requirements of ASR models. This technique allows the model to learn from a vast amount of decentralized data stored on user devices, thus effectively alleviating the privacy issue by ensuring raw data never leaves the user device~\cite{mcmahan2017communication,Zhao2018FederatedLW, karimireddy2020scaffold}. Recent studies demonstrated promising results with federated learned ASR models ~\cite{guliani2022enabling,lin2022federated}. Despite its merits, federated learning imposes both heavy computation burdens on participating clients and communication cost due to the frequent exchange of parameter updates between the server and clients during model training~\cite{wu2022communication, 9464278_short}. 
These issues become intensified by the rapid growth in parameter quantities in state-of-the-art ASR models, which increase from millions~\cite{gulati2020conformer} to billions~\cite{radford2022robust}. 

Besides, once an ASR model is trained and deployed in the real world, the service usually encounters problems like handling low-resource languages, dialects, accents, and registers~\cite{librilight}. There is a similar finding: ASR models' performance usually drops dramatically when the model is trained on a particular dataset while evaluated on another~\cite{52517}. Therefore, parameter-efficient domain adaptation presents a proper solution to handle the unique complexity of various scenarios.
When transferring the learning between different user domains, the method freezes the majority of a pre-trained model and only tunes a subset of components called adapter. The adapter tuning method is efficient and converges fast because only a tiny portion of parameters needs to be trained. The results show that when equipped with proper adapters, these models, only their adapter parameters are updated, can achieve comparable performance to their counterparts whose all parameters are tuned~\cite{52517, he2022towards, houlsby2019parameter, lowrank}. 

This paper studies the optimal strategy for domain adaptation using the adapter tuning method under federated learning. We alleviate the computation and communication cost of federated learning and provide a vast amount of on-device data for ASR model training. Our work is a natural extension of the domain adaptation with adapter tuning ~\cite{52517} under the new federated learning setting by examining the integration method of adapters into pre-trained models and designing of efficient adapters. Our key contributions include (1) A comprehensive analysis of various adapter efficiencies within federated learning, (2) The provision of an optimal solution for integrating adapters into pre-trained models during federated tuning, and (3) Evidence that federated adapter tuning can match the performance of centralized adapter tuning.

%% file: sections/2-method.tex
\section{Method}


To study the parameter-efficient domain adaptation under the federated learning setting, we design method from two perspective, how to incorporate adapters into models (training pipeline) and which adapter is more efficient (adapter study).

\begin{figure}
    \centering
    \includegraphics[width=1.0\linewidth]{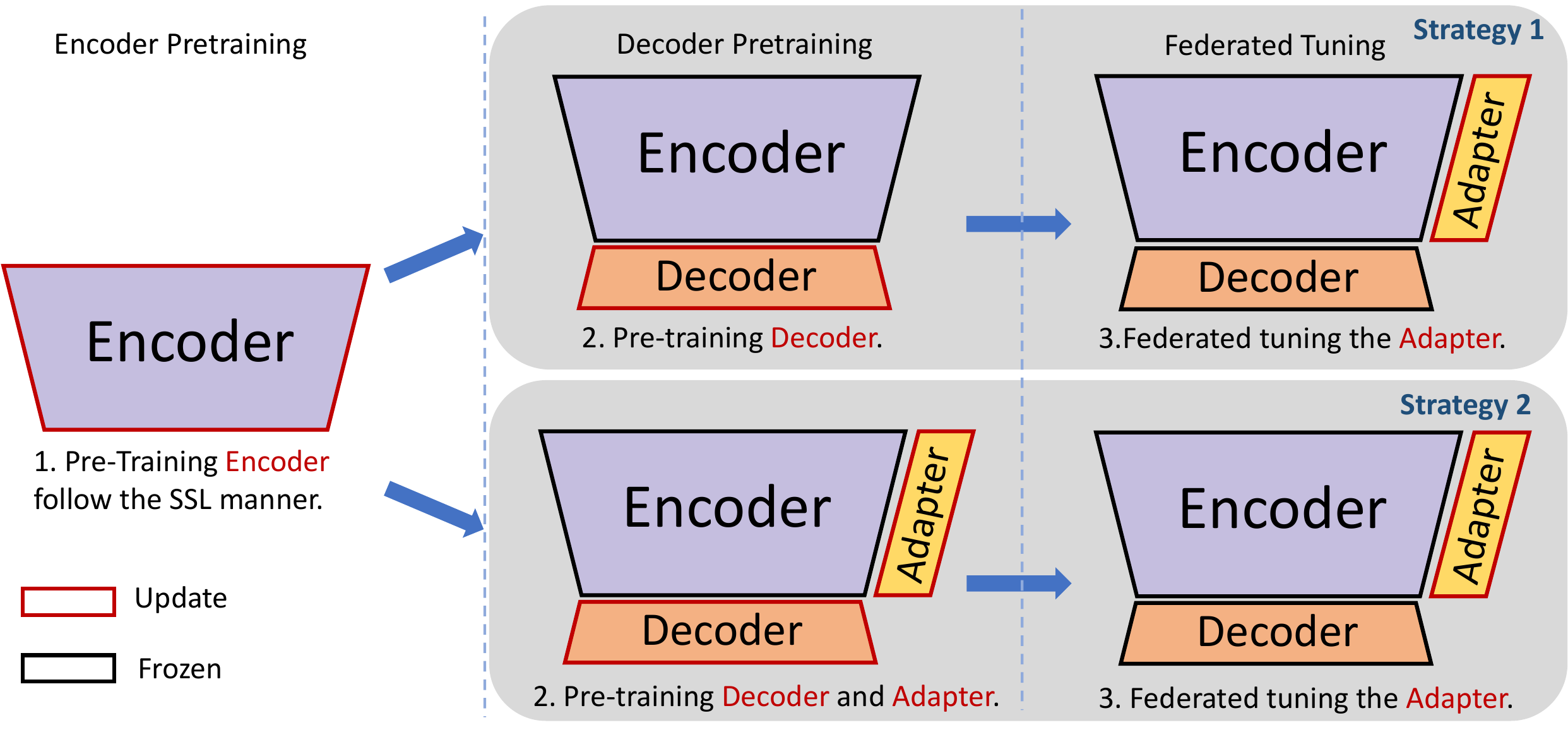}
    \caption{The pipeline incorporates 3 stages with 2 strategies.}
    \label{fig:training_pipeline}
\end{figure}

\begin{figure*}
    \centering
    \includegraphics[width=0.9\linewidth]{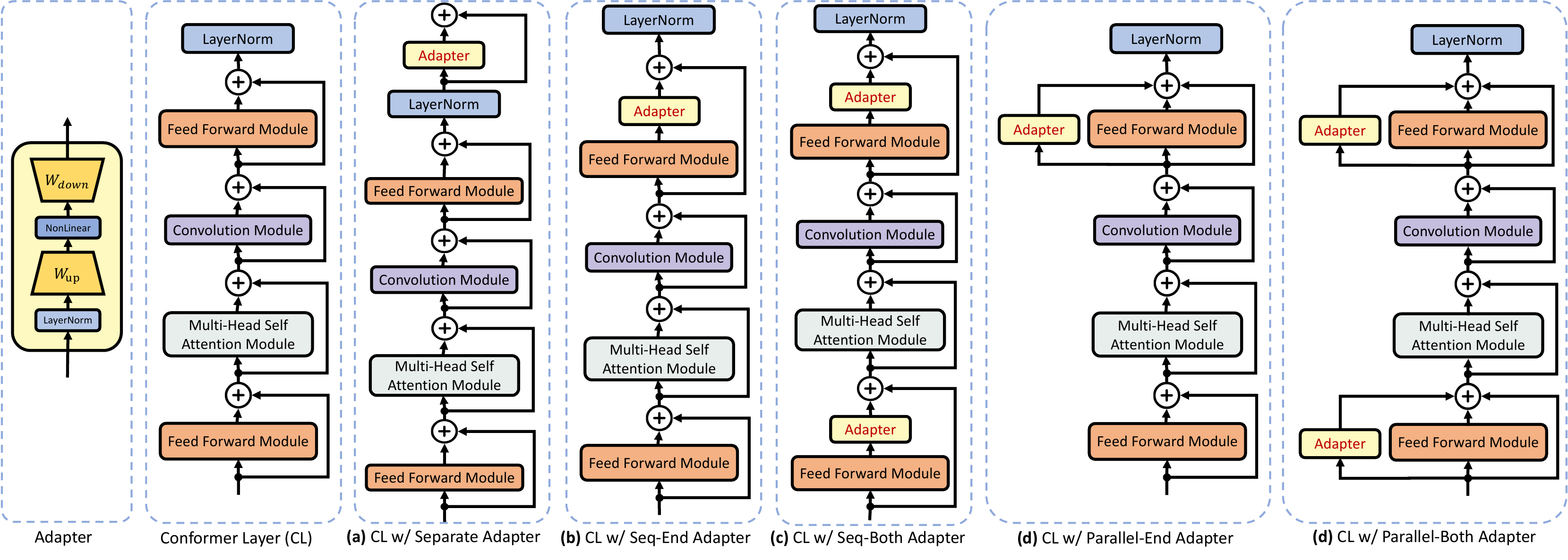}
    \vspace{-3ex}
    \caption{The structure of Adapter, Conformer Layer, and Conformer Layer w/ various Adapters.}
 
    \label{fig:adapter}
    \vspace{-2ex}
\end{figure*}

\subsection{Training Pipeline}
\label{sec:training_pipeline}
As detailed in Figure \ref{fig:training_pipeline}, our training pipeline incorporates three main stages. 

\noindent \textbf{Encoder Pretraining.} Following the self-supervised learning (SSL) method proposed by~\cite{pmlr-v162-chiu22a}, we employ a large and unlabeled dataset to pre-train the encoder, setting a foundation for the model parameters. During the SSL training, a segment of masked speech signals is transformed into a discrete label through a random projection, and the encoder is learned by predicting the discrete label of masked signals. Encoder pre-training aims to generate a powerful encoder that can convert speech signals to informative hidden representations.

\noindent \textbf{Decoder Pretraining.} We investigate two settings in this stage. The first one couples the pre-trained encoder with a decoder, keeping the encoder parameters static while utilizing a new dataset to train the decoder. The second setting equips the encoder with adapters, then freeze the encoder parameters and simultaneously trains adapters and the decoder.

\noindent \textbf{Federated Tuning.}  In this stage, we apply federated adapter tuning to examine the domain adaptation ability of different adapters under the federated learning context. The strategy slightly differs depending on whether these adapters exist in the previous stage. When adapters are missing in the pre-trained model, we incorporate adapters into the encoder, freeze both the encoder and decoder parameters, and then apply adapter tuning. Otherwise, we directly hold the encoder and decoder parameters static and tune adapter parameters.

\subsection{Adapter Design}

This section demonstrates the designs of various adapters in our work. The details on the adapter structure and how they are injected into each encoder layer are shown in Figure \ref{fig:adapter}.

\noindent \textbf{Adaptor Structure}. Our approach follows the adapter structure proposed in~\cite{houlsby2019parameter}. When given the hidden representation $\boldsymbol{h}$ from the previous layer, the adapter function $\boldsymbol{f}_{A}$ alters the representation as follows:
\begin{equation}
\boldsymbol{f}_{A}(\boldsymbol{h}) = \sigma\left(\boldsymbol{h} \boldsymbol{W}_{\text {down}}\right) \boldsymbol{W}_{\text {up}}
\end{equation}
where $\boldsymbol{W}_{\text {down}}$ and $\boldsymbol{W}_{\text {up}}$ represent learnable parameters and $\sigma$ symbolizes a non-linear function. 

\noindent \textbf{Adaptor Position}. Since the backbone model used in our study is Conformer~\cite{gulati2020conformer}, whose structure can be found in Figure \ref{fig:adapter}, as suggested by \cite{bapna-firat-2019-simple} and \cite{he2022towards}, the adapter can be interjected either between each Conformer layer or added near the feed-forward modules (FFM) within each Conformer layer. Thus, we propose three options: (1) \textit{Separate}, where the adapter is inserted between Conformer layers; (2) \textit{End}, where the adapter is placed in the last FFM of a Conformer layer; (3) \textit{Both}, where adapters are incorporated at both the beginning and end of FFMs in Conformer layer. 

\noindent \textbf{Insertion Mode}. When the adaptor position is set as either \textit{Both} or \textit{End}, there is flexibility in how to merge the FFM output with the adapter output. In line with He's settings~\cite{he2022towards}, we offer two options: (1) \textit{Parallel}, where given the FFM input $\boldsymbol{x}$ and output $\boldsymbol{h}$, the adapter modifies the representation as follows: $\boldsymbol{h}_{\text{new}} =\boldsymbol{h}+\boldsymbol{f}_{A}(\boldsymbol{x})$; (2) \textit{Sequential (Seq)}, where given the FFM output $\boldsymbol{h}$, the adapter modifies the representation as: $\boldsymbol{h}_{\text{new}} =\boldsymbol{f}_{A}(\boldsymbol{h})$.

Finally, considering the design space offered by the Adaptor Position and Insertion Mode, we can derive five types of conformer layers with adapters: (a) Conformer Layer with Separate Adapter, (b) Conformer Layer with Seq-End Adapter, (c) Conformer Layer with Seq-Both Adapter, (d) Conformer Layer with Parallel-End Adapter, and (e) Conformer Layer with Parallel-Both Adapter. Detailed design specifications for all of them can be found in Figure \ref{fig:adapter}.

%% file: sections/3-exp.tex
\section{Experiments}

\subsection{Setting}


\noindent \textbf{Dataset.} This study trains and evaluates models on three speech recognition datasets. Firstly, \textit{LibriLight}, developed by \cite{librilight}, is a large-scale corpus designed for self-supervised learning and other semi-supervised tasks in automatic speech recognition. It comprises approximately 60k hours of English audio and is used for \textit{encoder pretraining}.
Secondly, \textit{LibriSpeech}, developed by \cite{librispeech}, is a comprehensive English speech dataset derived from audiobooks within the public domain via the LibriVox project. It offers around 1k hours of speech data divided into various subsets for different training, validation, and testing scenarios. We use this dataset for \textit{decoder pretraining}. 
Finally, \textit{Fleurs}~\cite{conneau2022fleurs}, a multilingual benchmark dataset, is leveraged for \textit{federated tuning }with its EN-US subset that consists of roughly 12 hours of English audio.

\noindent \textbf{Model Architecture.} Our model architecture is built based on the Conformer~\cite{gulati2020conformer}, incorporating a heavy encoder (103.05M) and a light decoder (3.91M). The encoder consists of 17 Conformer layers with a hidden dimension of 512 and a CNN kernel size of 32. On the other hand, the lightweight decoder comprises an Embedding Prediction Layer and a Joint Layer, both with a dimension of 640. During federated training, different types of adapters can be inserted into each Conformer layer in the encoder for federated tuning. The parameter number of each adapter module is 4.47M. 

\noindent \textbf{Federated Training Setting.} We implement the FedAVG~\cite{mcmahan2017communication}
training process based on the FedJax~\cite{fedjax2021} framework. For each round, each client sends its parameter delta to the central server, and the central server will update its parameters with the average of each model delta and then send the updated model back to each client. We establish a federated setup with 64 clients, each having a batch size of 10, and the training process is conducted over 1k rounds with one iteration per round, considering the limited training resource on clients and the round setting proposed by FedJax. The server learning rate is set to $2 \times 10^{-4}$ with Adam as the optimizer, while the client learning rate is $10^{-4}$, employing SGD as the optimizer.

\noindent \textbf{Metric.} To evaluate the performance of our model in Automatic Speech Recognition tasks, we employ Word Error Rate (WER) as our main evaluation metric. WER provides a comprehensive measure of accuracy by quantifying the alignment between the predicted transcription and the ground truth.

\begin{table}[H]
\centering
\small
\caption{Pre-Train Model Performance. The best results are in \textbf{bold}, and the second best results are \underline{underlined}.}
\label{tab:pt}
\resizebox{0.8\linewidth}{!}{
\begin{tabular}{c ccc}
\toprule
\multirow{2.5}{*}{Architecture} &\multicolumn{3}{c}{\bf Performance(WER)}\\
\cmidrule(lr){2-4} 
& {LibriSpeech } & { }  & {Fleurs EN-US}\\
\midrule
PT w/o Adapter&\textbf{1.94/4.11}/\textbf{2.03}/\textbf{4.36}& & 30.58/28.86\\
\midrule
PT w/ Separate Adapter&2.12/4.42/2.11/\underline{4.45}& &30.70/ 29.98\\
PT w/ Seq-End Adapter&2.06/4.56/2.26/4.80& &\textbf{26.92}/ \textbf{25.23}\\
PT w/ Seq-Both Adapter&2.00/4.49/2.12/4.54& &\underline{27.62}/\underline{26.76}\\
PT w/ Parallel-End&2.03/4.39/2.15/4.56& & 30.97/29.86\\
PT w/ Parallel-Both&\underline{1.96}/\underline{4.36}/\underline{2.06}/4.62& &37.26/ 36.61\\
\bottomrule
\end{tabular}
}
\end{table}

\subsection{Performance and Analysis}
\begin{table*}[!htbp]
\centering
\small
\caption{Model performance comparison after federated tuning. The best results are in \textbf{bold}.}
\label{tab:federated}
\resizebox{1,0\linewidth}{!}{
\begin{tabular}{cccc cc cc c cc}
\toprule
\multirow{2.5}{*}{Index} &\multirow{2.5}{*}{PT Model} & \multirow{2.5}{*}{Adapter} &\multirow{2.5}{*}{Updated Para \%} &\multicolumn{2}{c}{\bf Performance (WER)}& & \multicolumn{2}{c}{\bf Compared with PT}\\
\cmidrule(lr){5-6} \cmidrule(lr){8-9} 
&&&& {LibriSpeech} & {Fleurs EN-US}& & {LibriSpeech} & {Fleurs EN-US}\\
\midrule
1&PT w/o Adapter & Separate Adapter &4.01\%&\textbf{2.04}/\textbf{4.18}/\textbf{2.18}/\textbf{4.39} & 29.99/28.19 &&0.1$\color{green}{\uparrow}$/0.1$\color{green}{\uparrow}$/0.2$\color{green}{\uparrow}$/0.0$\color{green}{\uparrow}$ & -0.6$\color{red}{\downarrow}$/-0.7$\color{red}{\downarrow}$\\
2&PT w/o Adapter & Seq-End Adapter&4.01\% &2.38/4.82/2.47/4.90 & 29.08/27.18&&0.4$\color{green}{\uparrow}$/0.7$\color{green}{\uparrow}$/0.4$\color{green}{\uparrow}$/0.5$\color{green}{\uparrow}$ & -1.5$\color{red}{\downarrow}$/-1.7$\color{red}{\downarrow}$\\
3&PT w/o Adapter & Seq-Both Adapter&7.71\% &2.73/5.33/2.82/5.36 & 28.60/27.11 &&0.8$\color{green}{\uparrow}$/1.2$\color{green}{\uparrow}$/0.8$\color{green}{\uparrow}$/1.0$\color{green}{\uparrow}$ & -2.0$\color{red}{\downarrow}$/-1.8$\color{red}{\downarrow}$\\
4&PT w/o Adapter & Parallel-End Adapter&4.01\% &2.23/4.58/2.31/4.76 & 29.44/27.57&&0.3$\color{green}{\uparrow}$/0.5$\color{green}{\uparrow}$/0.3$\color{green}{\uparrow}$/0.4$\color{green}{\uparrow}$ & -1.1$\color{red}{\downarrow}$/-1.3$\color{red}{\downarrow}$\\
5&PT w/o Adapter & Parallel-Both Adapter&7.71\% &2.45/4.95/2.50/5.04 & 28.56/26.88&&0.5$\color{green}{\uparrow}$/0.8$\color{green}{\uparrow}$/0.5$\color{green}{\uparrow}$/0.7$\color{green}{\uparrow}$ & -2.0$\color{red}{\downarrow}$/-2.0$\color{red}{\downarrow}$\\
\midrule
6&PT w/ Separate Adapter & Separate Adapter&4.01\% &2.19/4.54/2.25/4.61 & 30.26/29.39&&0.1$\color{green}{\uparrow}$/0.1$\color{green}{\uparrow}$/0.1$\color{green}{\uparrow}$/0.2$\color{green}{\uparrow}$ & -0.4$\color{red}{\downarrow}$/-0.6$\color{red}{\downarrow}$\\
7&PT w/ Seq-End Adapter & Seq-End Adapter&4.01\% &2.18/4.74/2.36/5.05 & \textbf{26.79}/\textbf{25.20}&&0.1$\color{green}{\uparrow}$/0.2$\color{green}{\uparrow}$/0.1$\color{green}{\uparrow}$/0.2$\color{green}{\uparrow}$ & -0.1$\color{red}{\downarrow}$/-0.0$\color{red}{\downarrow}$\\
8&PT w/ Seq-Both Adapter & Seq-Both Adapter&7.71\% &2.22/4.85/2.29/4.93 & 27.49/26.75 &&0.2$\color{green}{\uparrow}$/0.4$\color{green}{\uparrow}$/0.2$\color{green}{\uparrow}$/0.4$\color{green}{\uparrow}$ & -0.1$\color{red}{\downarrow}$/-0.0$\color{red}{\downarrow}$\\
9&PT w/ Parallel-End Adapter & Parallel-End Adapter&4.01\% &2.22/4.61/2.39/4.76 &30.22/29.52&&0.2$\color{green}{\uparrow}$/0.2$\color{green}{\uparrow}$/0.2$\color{green}{\uparrow}$/0.2$\color{green}{\uparrow}$ & -0.8$\color{red}{\downarrow}$/-0.3$\color{red}{\downarrow}$\\
10&PT w/ Parallel-Both & Parallel-Both Adapter&7.71\% &2.19/4.85/2.29/4.97 & 36.84/36.05 &&0.2$\color{green}{\uparrow}$/0.5$\color{green}{\uparrow}$/0.2$\color{green}{\uparrow}$/0.3$\color{green}{\uparrow}$ & -0.4$\color{red}{\downarrow}$/-0.6$\color{red}{\downarrow}$\\
\bottomrule
\end{tabular}
}
\vspace{-3ex}
\end{table*}

\begin{figure}[H]
    \centering
    \includegraphics[width=0.8\linewidth]{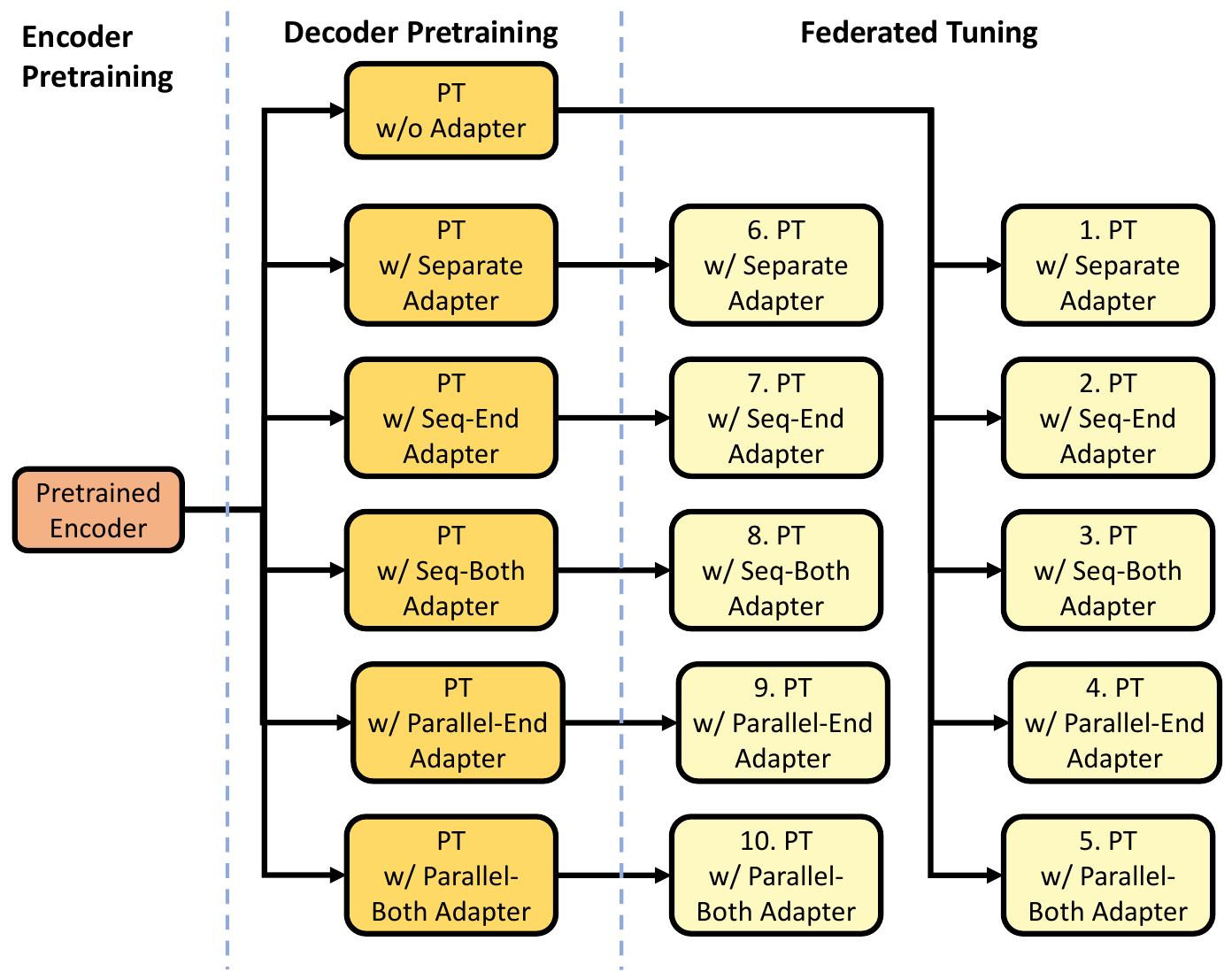}
    \caption{The model family of experiments. Beginning with a Pre-trained Encoder in the first stage, we extend to 6 different Pre-Trained models in the second stage after integration with various adapters (model performances are summarized in Table \ref{tab:pt}), and ultimately derive 10 models with unique Pre-Trained bases or adapters during the federated tuning process (model performances are summarized in Table \ref{tab:federated}, the model index in Figure is matched with the index column in Table \ref{tab:federated}).}
    \label{fig:model_family}
\end{figure}


As shown in Section \ref{sec:training_pipeline}, our model training is divided into 3 stages, each contributing to the development of the model family as depicted in Figure \ref{fig:model_family}. In the 1st stage, the LibriLight dataset is used to pre-train the encoder, and the chosen encoder is at the 100k steps. For the 2nd stage, 6 models' decoders and adapters are trained using the LibriSpeech dataset, and all models halt at the 80k steps. After the 3rd stage, 10 models are obtained after 1k steps of federated tuning.

\noindent \textbf{Pre-Trained Model Performance.} Table \ref{tab:pt} consolidates all 6 pre-trained model performance. This table shows that without adapters, the model achieves optimal performance on the decoder training dataset, LibriSpeech. This could be due to the potential destabilization of the encoder output when equipped with adapters, making learning a superior decoder challenging. If equipped with adapters, (1) When having the same number of tuning parameters, the transfer ability is Parallel $>$ Sequential $>$ Separate; (2) More tuning parameters usually indicate better transfer ability. This result is aligned with the findings in~\cite{he2022towards}. Besides, better transfer ability usually means loss of generalization ability from the original pre-train model (worse performance in Fleurs EN-US).

\noindent \textbf{Federated Tuning Model Performance.} In the federated tuning phase, 10 models are generated, and their performances are presented in Table \ref{tab:federated}. Performance patterns indicate a trade-off between the pretraining dataset, LibriSpeech, and the user-specific dataset, Fleurs EN-US; as performance on the former deteriorates, it improves on the latter. The PT w/ Separate Adapter, based on the PT w/o Adapter, exhibits optimal LibriSpeech performance due to its superior pre-trained model and the weak transfer ability of the Separate Adapter. Meanwhile, for the performance on the Fleurs EN-US dataset, although all model performances improve after federated tuning, the gain is relatively small compared with the performance gap among different pre-trained models.


We summarize our result here: (1) Adapter behaviors in domain adaptation present similar trends in centralized and federated learning. Parallel adapters generally outperform Sequential and Sequential surpassing Separate adapters. Also, adapters with more parameters show better transfer learning ability than those with fewer parameters; (2) From the generalization perspective, we identify a trade-off between adapting to a new domain and preserving performance on the original domain, highlighting that highly parameter-efficient adapters usually risk compromising performance in the original domain; (3) Adapters can save a lot of communication burden and computation resources, reducing the updated parameters from 106.96M to 8.96M; (4) A powerful pre-trained model is necessary to delivery high-quality ASR service.

\begin{figure}[htbp]
     \centering
    \includegraphics[width=\linewidth]{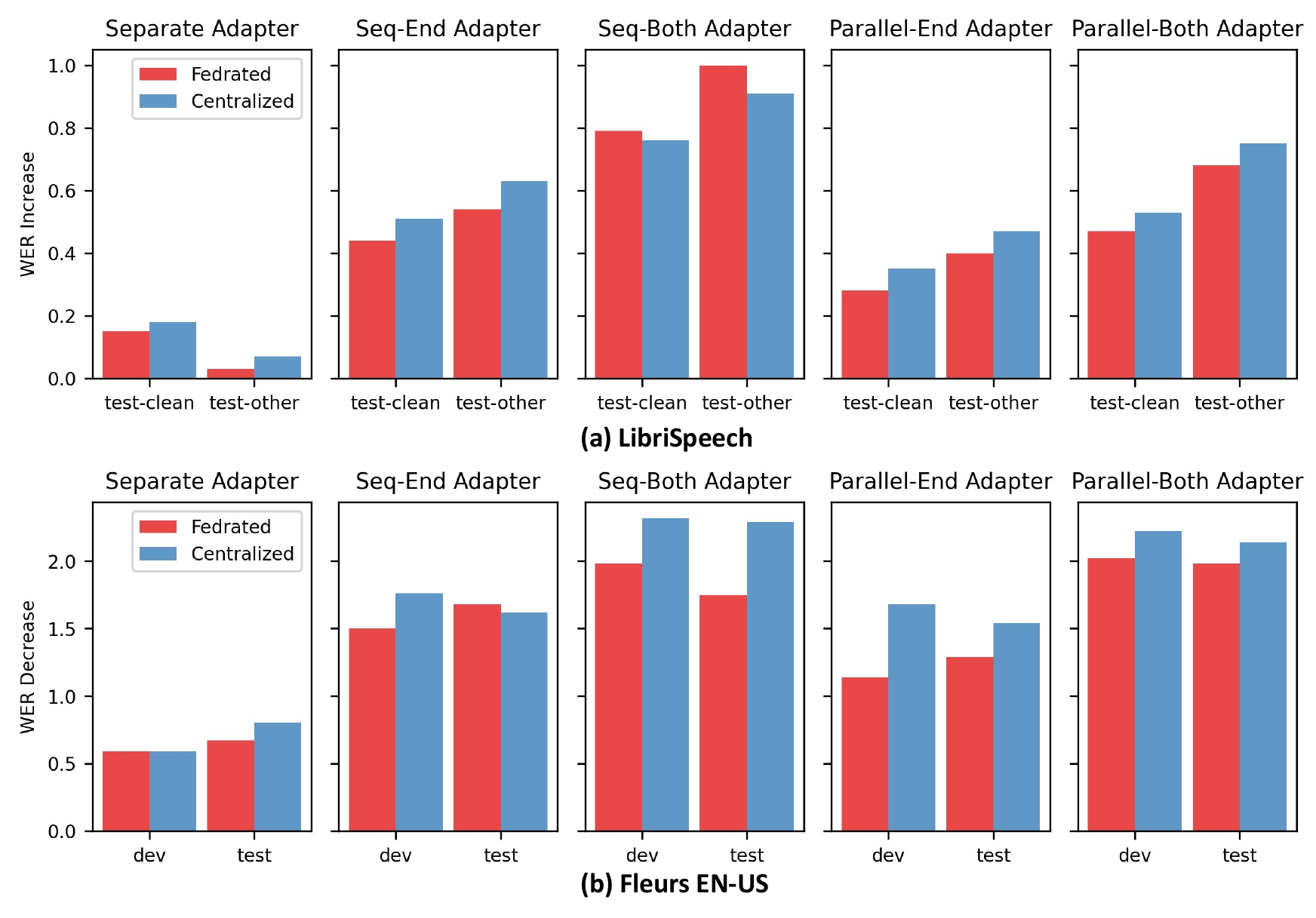}
    \vspace{-5ex}
    \caption{The WER change when tuning PT w/o Adapter model on the Fleurs EN-US dataset. Each column represents a model equipped with a different adapter; the first row depicts the increase in WER for the LibriSpeech dataset post-tuning, while the second row shows the corresponding decrease in WER for the Fleurs EN-US dataset.}
    \label{fig:fed_expriment}
    \vspace{-3ex}
\end{figure}

\subsection{Ablation Study}

This section compares our federated training models with centralized training models. Using the PT w/o Adapter model as a base, we apply both federated and centralized training, incorporating various adapters, on the Fleurs EN-US dataset, ensuring an equal number of training samples for each by setting 5k iterations with a batch size of 128 for centralized training. As illustrated in Figure \ref{fig:fed_expriment}, the performance of centralized training models usually shows a larger increase/decrease compared to their federated counterparts, though the overall performance between the two settings is similar. However, when tuning PT w/ Seq-Both Adapter, the model performance decreases more rapidly under federated training, indicating that federated learning can occasionally be less stable than centralized training.

%% file: sections/4-con.tex
\section{Conclusion}

In this work, we investigate the potential of applying domain adaption with federated learning to improve ASR models under user-specific scenarios and present a detailed study of various adapters and strategies for this setting. Finally, we show that federated adapter tuning could match the performance of a centralized counterpart, paving the way for future research.